\documentstyle[ijmpc,epsfig]{article}
\def\op{{\bf D}}
\def\dm{{\delta m^2}}

\def\rhs{{\bf f}}
\def\e{{\bf e}}
\def\de{{\delta \e}}
\def\f{{\bf f}}
\def\r{{\bf r}}
\def\S{{\bf S}}

\def\matr#1{{\bf #1}}
\def\SUZWEI{{\rm SU(2)}}
\newcommand{\A}{{\cal A}}
\newcommand{\calH}{{\cal H}}
\newcommand{\dA}{{\delta\A}}
\newcommand{\C}{{\cal C}}
\def\tilde#1{\widetilde{#1}}
\def\xitilde{{\tilde{{\xi}}}}
\newcommand{\exceq}{\buildrel ! \over =}
\title{Another Look at Neural Multigrid}
\author{Martin B\"aker}
\def\qed{\hbox{${\vcenter{\vbox{                        %HOLLOW SQUARE
   \hrule height 0.4pt\hbox{\vrule width 0.4pt height 6pt
   \kern5pt\vrule width 0.4pt}\hrule height 0.4pt}}}$}}

\renewcommand{\thefootnote}{\fnsymbol{footnote}}        %USE SYMBOLIC FOOTNOTE

\def\bsc{{\sc a\kern-6.4pt\sc a\kern-6.4pt\sc a}}       %LATEX LOGO
\def\bflatex{\bf L\kern-.30em\raise.3ex\hbox{\bsc}\kern-.14em 
T\kern-.1667em\lower.7ex\hbox{E}\kern-.125em X} 

\begin{document}
\runninghead{Another Look at Neural Multigrid}{Another Look at Neural
  Multigrid} 

\normalsize\textlineskip
\thispagestyle{empty}
\setcounter{page}{1}

\copyrightheading{}                     %{Vol. 0, No. 0 (1993) 000--000}

\vspace*{0.88truein}

\fpage{1}
\centerline{\bf ANOTHER LOOK AT NEURAL MULTIGRID}
\title{ANOTHER LOOK AT NEURAL MULTIGRID}

\author{Martin B\"aker}
\vspace*{0.37truein}
\centerline{\footnotesize Martin B\"aker\footnote{e-mail:
    $<$baeker@x4u2.desy.de$>$}}
\vspace*{0.015truein}
\centerline{\footnotesize\it II. Institut f\"ur
    Theoretische Physik der Universit\"at Hamburg, Luruper Chaussee 149}
\baselineskip=10pt
\centerline{\footnotesize\it 22761 Hamburg , Germany}
\vspace*{0.225truein}
\publisher{(received date)}{(revised date)}

\vspace*{0.21truein} \abstracts{ We present a new multigrid method
  called neural multigrid which is based on joining multigrid ideas
  with concepts from neural nets. The main idea is to use the
  Greenbaum criterion as a cost functional for the neural net.  The
  algorithm is able to learn efficient interpolation operators in the
  case of the ordered Laplace equation with only a very small critical
  slowing down and with a surprisingly small amount of work comparable
  to that of a Conjugate Gradient solver.  }{In the case of the
  two-dimensional Laplace equation with $\SUZWEI$ gauge fields at
  $\beta=0$ the
  learning exhibits critical slowing down with an exponent of about
  $z\approx0.4$. The algorithm is able to find quite good
  interpolation operators in this case as well. Thereby it is proven
  that a practical true multigrid algorithm exists even for a gauge
  theory. An improved algorithm using dynamical blocks tthat will hopefully
  overcome the critical slowing down completely is sketched. }{}

\vspace*{10pt}
\keywords{Discretized differential equations, multigrid, neural nets,
  disordered systems, lattice gauge theory}
\vspace*{1pt}\textlineskip      
\setcounter{footnote}{0}
\renewcommand{\thefootnote}{\alph{footnote}}

\section{Introduction}\noindent
Multigrid methods are among the most successful strategies for 
solving discretized differential equations. In the presence of
disorder, which is the case of most interest to contemporary physics,
the development of methods that are able to deal with the lost
translational invariance of the system has proven extremely
difficult. As an example we may look at the efforts to find a
multigrid solver for the Dirac equation in Lattice Gauge Theory,
reviewed in\cite{ThomasReview}. Up to now, no true multigrid algorithm
has been found that is able to deal with this problem as efficiently
as we would like, although there is a {\em unigrid\/} method called
ISU that has been proven to work in two dimensions\cite{ISU}.

The development of the ISU algorithm has been triggered by an attempt
to connect multiscale ideas with the method of neural nets
\cite{NMG}. A more direct combination of multigrid and neural net
methods was based on the idea 
to learn the shapes of slow-converging modes by a standard
back-propagation scheme, using the known errors of test problems as
targets. This, however, was shown to be not feasible \cite{Bernhard}.
The main reason was that the cost functional possessed only a very
narrow minimum, which was difficult to find by the neural
net whenever the criticality of the problem operator (i.e.\ its
condition number) was large.

In this work we retain the idea of using test problems to do the
learning, but we do not try to learn them as in a pattern
recognition neural network. Instead of this, we borrow ideas from the
ISU algorithm\cite{ISU} and the principle of indirect
elimination\cite{PIE}.

\section{The problem}\noindent
Consider a linear operator~$\op$ which may arise from a discretized
differential equation defined on a cubic
lattice~$\Lambda^0$.\footnote{It is not necessary to restrict our
  attention to the case of a cubic lattice, but it eases the
  implementation of the method.}\ Here and in the following we
assume~$\op$ to be {\em positive definite}, if it were not, we could
use the operator~$\op^\ast \op$ instead. The general form of the
equation to be solved is then
\begin{equation}
  \label{fundamental_eq}
  \op \xi = \rhs \quad .
\end{equation}
It is well known that standard solvers like Conjugate Gradient or
Overrelaxation show the phenomenon of critical slowing down: The
number of iterations needed to solve the equation with a given
precision scales with some power of the condition number (quotient of
the largest and smallest eigenvalue). 

At each time-step, any iterative method  will yield an {\em approximate
solution\/}~$\xitilde$. 
We introduce two important quantities: the {\em error\/}~$\e= \xi
-\xitilde$ which is the difference between the true and the actual
solution and is of course not known, and the {\em residual\/}~$\r= \f
- \op \xitilde$, the difference between the true and the actual
righthandside. With these definitions we can recast the fundamental
equation~(\ref{fundamental_eq}) as
\begin{equation} \op\, \e = \r \quad, \label{error_eq} \end{equation}
called the {\em error equation}. 

For a linear method, we can also introduce the {\em iteration matrix}~$\S$
which tells us what the new error after the next iteration step will
be, given the old one:
\begin{equation} \e^{\rm new} = \matr{S}\, \e^{\rm old}
  \quad. \label{IterError} \end{equation}

\section{The Neural Multigrid}
Although our method is motivated by ideas borrowed from neural network
methods, a thorough understanding of these methods is not necessary. A
good introductory text is \cite{NeuralNets}. For introductions to
multigrid see \cite{Brandt,Briggs,Hack}.

The basic setup of a multigrid algorithm uses {\em auxiliary
  lattices}, also called {\em block lattices\/} or {\em coarse grids},
$\Lambda^1,\Lambda^2, \ldots,\Lambda^N$ with lattice spacings~$a_j
=L_b^j a_0$, where $L_b$ is the blocking factor and is usually chosen
to be~$2$. 
The last lattice~$\Lambda^N$ should contain few enough points that
a direct solution of any equation living on this lattice is easy.

Let $\calH^j$ be the space of functions on lattice $\Lambda^j$. Then
we introduce {\em grid transfer operators:}
\begin{eqnarray} {\rm the\ interpolation\ operators:\quad }&
                               \A^k &: \calH^{k+1} \mapsto \calH^k
{\quad \rm and} \\
 {\rm the\ restriction\ operators:\quad }&
                               \C^{k} &: \calH^k \mapsto \calH^{k+1}
{\quad ,}\end{eqnarray}
with $k<N$. So operators $\A^{k}$ interpolate from a coarser
to a finer grid, whereas the restriction operators do the reverse. 
For reasons of efficiency these operators are not allowed to
interpolate from one block grid point to all of the fine grid: if we
identify a block grid point~$x$ with its corresponding fine grid point~$z$
we must require $\A^k(z,x) = 0$ unless $x$ lies near~$z$.
We
will always choose the operators to be adjoints of each other:
$\C^{k} = {\A^k}^\ast$. We also recursively define {\em effective
  operators\/} $\op^k = \C^{k-1} \op^{k-1} \A^{k-1}$, of course
setting $\op^0 = \op$ to stop the recursion.

Multigrid methods are based on the observation that standard
relaxation al\-go\-rithms\cite{Varga,Young} usually smoothen the
error, or in more general terms, project onto the lower half of the
eigenspectrum of the problem operator~$\op^0$.  If we can find
interpolation operators $\A^0$ such that the smooth error lies within
their range, we can write $\e^0 \approx \A^0 \e^1$. Inserting this
into the error equation~(\ref{error_eq}) we find
\begin{eqnarray}
    \op^0 \, \A^0 \, \e^1& =& \r^0\\
 \C^0\, \op^0 \,\A^0\, \e^1& =&\C^0\, \r^0 \\
    \op^1 \,\e^1& =& \r^1 \label{block_eq}\quad,
\end{eqnarray}
involving only quantities on the block lattice. Solving this equation
and interpolating back yields a good estimator for the
error~$\e^0$. The block-equation~(\ref{block_eq}) can be solved
recursively by going to a still coarser grid, until we reach the
coarsest layer $\Lambda^N$ where the equation can be solved directly.

After this review of multigrid methods let us now set up the neural multigrid.
The basic fact we are using is the Greenbaum
criterion\cite{Greenbaum}, which up to now has been considered to have
no practical use at all. It states that optimal convergence of the
multigrid is achieved when the interpolation operators are able
to represent all the modes of the system that are slowly converging
under the used relaxation process. In other words, all the highest
eigenmodes of the iteration matrix of the relaxation process have to
lie within the range of the interpolation operators. Stated like this it
is clear why the practical value of this principle is small: How
should we know all of the bad-converging modes of the system?

To learn these modes we are going to use {\em test problems}: A test
problem is a problem with a known solution, chosen such that the
initial error is a random function on the lattice. This can be done by
drawing the exact solution $\xi$ from a random distribution, calculating
the test problem's righthandside $\f = \op \xi$ and starting with
the initial guess $\xitilde=0$.

A first idea to exploit the criterion uses the fact that indeed we
have a good projector onto the bad-converging modes, namely the
relaxation process itself. Therefore it would be possible to start
with a test problem, do some relaxations and thereby project the
initial error onto the space of the slow-converging modes. This error
could then be learned (in some way) by the neural multigrid.

Of course this method is doomed to fail: To exhaust the complete space
of the slow modes would take a very long time because the projection
on the worst modes will only be efficient after many relaxation
sweeps. This means that the projection method itself suffers from
critical slowing down, and therefore our neural multigrid will as
well.

Nevertheless, our new method is based on this idea, but with another
ingredient, which is similar to the spirit of ISU: After a few
relaxation steps, the remaining error is something that should
definitely be learned. So we adapt our interpolation
operators to this error (the details will be explained later on), and
then {\em use the newly learned interpolation operators to further
  reduce the error.\/} To do so, we can make a coarse-grid correction
step. (For the time being, think of the method as a twogrid-method,
i.e.\ $N=1$.)
This then efficiently removes all those error components already
learned. We then relax again and start the learning process anew (of
course using everything we have already learned). 

It is easy to see why this avoids the pitfall described above: After
the first relaxation step the error will contain contributions from
all slow modes. Some of these will be learned by the multigrid and are
therefore removed. In this way we successively project out everything
that is not yet learned and then learn it. After each learning step,
we measure the convergence rate of the algorithm as it stands now by
solving {\em another\/} test problem. If this is sufficient (for
instance, if the error is reduced by a factor of~2 during each
multigrid cycle), we stop the learning procedure, otherwise we
continue.

It may happen that one coarse-grid correction step is not enough to
project the error onto the nullspace of the interpolation operators.  
In this case we may use the error~$\e_M$ of the measurement iteration
for the next learning step by setting $\f = \op \e_M$. This was the
method actually used to obtain the results shown below.

%Although this sounds like a
%nice idea, it is not a priori clear that this method will ever
%finish. Who guarantees that it is possible to learn all the slow modes
%and that we will not fall into an endless cycle, first learning some
%modes, then some others, but thereby forgetting about the
%first-learned modes, then learning these again, and so on? 
%Obviously, no one can assure that the method will work for every
%problem we pose to it.
%However, we believe that the method will work whenever a multigrid
%treatment of the problem is possible at all. If it is possible to have
%all slow modes within the range of the interpolation operators, then
%the method should be able to learn these operators. Whether this is
%true has to be proven in each case by a careful numerical study.

\section{The learning process}
Let us now look at the method in greater detail: How is the learning
actually done?

The first thing in setting up a neural network is to decide on a cost
functional that decides how good a certain network configuration
is. As we want to have the error after relaxation within the range of
the interpolation operators, we choose as functional
\begin{equation} E = \frac{\|\A \,\de - \e\|^2}{\|\e\|^2}
  \quad,\end{equation}
  where $\A$ is the interpolation operator, $\e$ is the actual error
  and $\de$ is chosen such as to minimize $E$ with the given
  interpolation operators, at least approximately. $\de$ is the
  coarse-grid correction we would get using these interpolation
  operators, so it is a function living on the coarse grid. With $x$
  being a coarse grid point and $z$ a fine grid point, we can rewrite
  $E$ using indices as 
\begin{equation} E = \frac{\sum_z\left(\sum_x\A(z,x) \,\de(x) -
    \e(z)\right)^2}{\sum_z\e(z)^2}
  \quad.\end{equation}
  
There are two possibilities to use this cost functional: The standard
way in neural networks would be to calculate the derivatives $\partial
E / \partial\A(z,x)$ and do an adjustment step in direction of this
gradient, perhaps also including some momentum term \cite{NeuralNets}. 

However, we can do better than this: Remembering that our
interpolation operators are localized objects, we see that at each
point the derivative only depends on very few of the interpolation
operators. We can therefore minimize~$E$ {\em exactly\/} at each point
so that the actual error lies within the range of the interpolation
operators. To do so, we require $\partial E / \partial\A \exceq 0$ at each
point. This is possible only because the space of 
interpolation operators is larger than the space of fine grid
functions, as most of the fine grid points are covered by more than
one interpolation operator. In fact, the system of equations
$\partial E / \partial\A \exceq 0$ is {\em under-determined\/} because of
this. This is an advantage as we can adjust the interpolation
operators and still retain some memory of their old values.

To describe the rules of the learning, let us define the
difference vector ${\bf d}(z) = \sum_x\A(z,x) \,\de(x) - \e(z)$, which
is nothing but the error after the correction step using $\de$. By
$\dA(z,x)$ we denote the change in $\A$ at the specified point. The
condition to be fulfilled is therefore that 
${\bf d}(z)$ would be zero after the change of~$\A$:
\begin{equation} \sum_x(\A(z,x)+\dA(z,x)) \,\de(x) - \e(z) \exceq
  0\quad.\label{constraint}\end{equation}
  As this system of equations is under-determined, we may add another
  requirement. In order to keep some memory of already learned things,
  we require that $\sum_{x,z} \dA(z,x)^2\exceq \min$, so we look for
  that choice of $\A$ that is closest to the old $\A$.

To solve these equations we can use the method of Lagrangian
multipliers (minimizing the change in~$\A$ and using
eq.~(\ref{constraint}) as a constraint) and we find
\begin{equation} \dA(z,x) = - \frac{{\bf d}(z)\, \de(x)}{\sum_{x^\prime,
      x^\prime \ni z} \de(x^\prime)} \quad.\end{equation}
Here the condition $x^\prime \ni z$ means that we have to use all
those points $x^\prime$ from which the interpolation operator can
reach the point~$z$. 

Finally, let us remark on the choice of $\de$: The obvious method to
determine it would be to use a full coarse-grid-correction
step. A somewhat simpler (and cheaper) choice is to choose $\de$ such
that the error is put to zero at those fine grid points only reached
by one interpolation operator, i.e.\ in the center of each block. This
should result in a good approximation of the optimal choice of $\de$
and is much cheaper. Up to now, this method has been used.

%The algorithm for the neural multigrid (in the twogrid case) is as follows:
%\begin{quote} THIS SHOULD BE ADAPTTED ACCORDING TO THE NEW METHOD!? OR
%  EVEN LEFT OUT?\\  
%\obeylines
%Set up the test problem
%Relax on~$\xi$ (presmoothing)
%Calculate~$\e$
%Adapt according to method described above
%Coarse-Grid-correct using the interpolation operators
%Relax on~$\xi$ (postsmoothing)
%Measure convergence rate
%If convergence rate is small enough, leave the learning process
%\end{quote}

\newpage
\section{The true multigrid}
For a true multigrid, all we have to do is to use the algorithm above
recursively. Whenever a coarse-grid-correction is required, we try to
solve the coarse grid equation, monitoring the convergence. If the
convergence is not good enough, we set up a test problem on the coarse
layer, learn good interpolation operators on this layer and then solve
the coarse grid equation again.

This looks like a very expensive method, for the average number of
learning cycles required on a certain layer will quickly blow up the
coarser the layer is. If we need only two learning steps in each
learning process the method will proceed through the multigrid in a
W-cycle fashion --- more learning  steps are analogous to higher
cycles. In two dimensions, the maximal cycle index allowed is four, so
we will quickly reach a limiting point where learning becomes
extremely slow and has some kind of critical slowing down.

Two facts oppose this tendency for the process to get slow: First of
all, as long as the interpolation operators are not yet very good, the
coarse grid equation will not be as critical as the fine grid
equation, so solving it is easier. Secondly, as the changes on each
layer are gradual, not every change on a fine layer will need
adaptations on coarser layers.

Of course, a definite answer can only be found by putting the method
to the test.

\section{Possible improvements}
\noindent\subsection{Adding an indirect elimination}\noindent
A very simple improvement that can be added to the algorithm is to
introduce an update based on the principle of indirect elimination: If
the convergence rate is not satisfactory, we store the error after a
measurement and do a line search in the direction of this error vector
before each multigrid cycle. As the error after the measurement
corresponds (at least approximately) to the worst-converging mode of
the algorithm, the line search will take care of this mode, so that it
will not contribute to the error in the following multigrid steps. A
thorough explanation of the principle of indirect elimination can be
found in \cite{PIE}. 

\subsection{Adding a memory}\noindent
Another possibility is to store some of the errors learned and to {\em
  relearn} them in later learning cycles. As we exactly minimize the
cost function in each step, we may forget some of the error shapes
learned earlier. If we store these errors, we can use them again. In
this way we attempt to find interpolation operators that are
able to deal with all the learned error shapes. It is well-known from
the neural net context that a shape once learned might be forgotten
later on if it is not shown again to the network. To be more precise,
what we add to the algorithm are the following steps:

Instead of the simple adaptation step we have
\begin{quote}
\obeylines
{\bf do \em nrOfMemorySteps \rm times:}
{\quad\em Adapt to the actual error}
{\quad\em Adapt to all memorized errors}
{\em Adapt to the actual error again}
\end{quote}
In practice, we usually choose to do three memory steps. We store only
those errors for later relearning whose learning has reduced the
convergence time appreciably.

\section{Results for the ordered case}\noindent
As a first test we investigated the neural multigrid for the standard
Laplace equation in two dimensions on a square lattice. We use
periodic boundary conditions and add a small mass term:
\begin{equation} (\Delta + m^2) \, \xi = f\quad.\end{equation}
It is well
known that in this case a standard multigrid method will exhibit
excellent convergence properties, reducing the error at least by a
factor of ten in each multigrid cycle\cite{Brandt,Hack}. 

We studied the behaviour of our neural multigrid on lattices of
size $16^2$ up to $64^2$. The coarsest layer was always chosen to
have a size of $2^2$, so that the scaling behaviour of the algorithm
could be studied. Remember that this is one of the key questions for
this method: How much do the additional learning steps on the coarser
layers affect the overall work done by the algorithm?

For a simple test case in one dimension it was found that the
situation is not favorable: The larger the lattice became, the
greater was the amount of work needed to learn good interpolation
operators. We estimated a critical exponent of 0.5. However, the
larger the dimension, the smaller is the relative size of the block
lattices compared to the finest lattice, so in higher dimensions the
situation might get better.
\begin{figure} \begin{center}
\epsfig{file=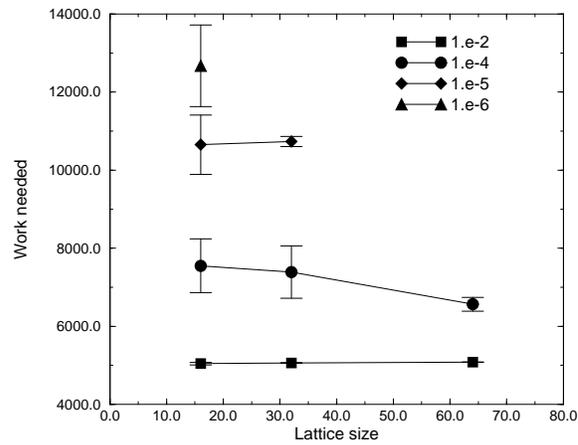,width=8cm}
\end{center}
\caption{Number of Work units (elementary vector operations; see text)
  needed by the neural multigrid to learn
  interpolation operators with a reduction rate of smaller
  than~0.1 and to solve the simple Laplace equation. For each data point, ten
  runs were made. The work does not depend on the lattice size, but
  for very small eigenvalues the needed work increases. (At $10^{-6}$
  one of the runs on the $16^2$-lattice did not converge, this was not
  taken into account here, see text.) Note the offset of the y-axis.\label{trivlapwork}}
\end{figure}

Figure~\ref{trivlapwork} shows the work needed to learn interpolation operators that
achieve the textbook efficiency of a reduction factor of at least~10 in each
multigrid cycle for the two-dimensional case. The work was measured in elementary vector
operations, i.e.\ each vector addition, multiplication on the finest
layer etc.\ counts as one work unit, a vector operation on the first
block layer as $1/4$ work units and so on. For each set of parameters
ten runs have been made. (Note that although the problem operator is always
the same, the righthandside and the randomly chosen start vectors
differ on each run.)

We can see two things: The number of work units needed does not grow
with the lattice size. This is an encouraging result because it means
that the recursive structure of the method does not lead to an
impractical algorithm. 

On the other hand we can see that the needed work does increase with
the criticality of the problem. For mass values smaller than $10^{-4}$
the work grows appreciably when we lower the mass further. We see that
the work needed at masses $10^{-5}$ is approximately twice that we
need at mass $10^{-2}$; out of this we would estimate a critical
exponent of $0.2$. (The critical exponent~$z$ for fixed lattice size
is given by $\hbox{\em(Work needed)} \propto \kappa^{z/2}$, where $\kappa$
is the condition number.) This is a fairly small value (e.g.~when
compared to the value for Conjugate Gradient algorithms with
$z=1$). Of course, with the present data we cannot exclude that the
work needed will grow faster when we decrease the mass even further.

At a mass value of $10^{-6}$ on a $16^2$-lattice, the neural multigrid
did not reach the required convergence rate within 20~learning cycles
in one of the ten runs. However, it did achieve a reduction factor
of~7 within a few iterations. This run has not been taken into account
in the picture, but it shows again that the algorithm has greater
difficulties at very small eigenvalues.

\begin{figure} \begin{center}
\epsfig{file=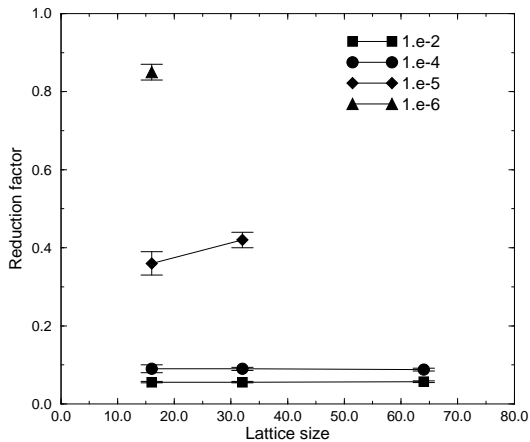,width=8cm}
\end{center}
\caption{Reduction factor~$\rho$ achieved by the neural multigrid after
  one learning step on the finest lattice has been completed. Again
  there is no dependence on the lattice size, but for very small mass
  values of the Laplace equation the reduction factor is gets large.\label{trivlaprho}}
\end{figure}

We can also look at the reduction rate~$\rho$ that is achieved after
the first learning step on the finest layer is finished, see
figure~\ref{trivlaprho}. Again we see that for masses of $10^{-4}$ and
larger the results are very good and textbook efficiency is usually
achieved; for smaller mass values the convergence deteriorates.

Although the results are not too bad, we can overcome this deterioration
by adding an improvement as explained in section~6.1. The
application of the principle of indirect elimination will eliminate
the contribution from the lowest eigenmode of the problem operator,
which is the one causing the problems here. After adding such an
updating step the convergence stays constant regardless of the mass
value. Note, however, that this is only possible because the number of
low-lying eigenmodes is small in the case of the Laplace equation.

%How do the interpolation operators look like? The standard
%interpolation operators used for ordered problems, written in stencil
%notation, are 
%\begin{equation} \A = \left[\matrix{1/4&1/2&1/4\cr
%1/2&1&1/2\cr1/4&1/2&1/4\cr}\right]\quad.\label{IOstencil}\end{equation} 
%The interpolation operators learned by the neural multigrid are always
%very close to these values and in all cases fulfills the 
%condition $\sum_x \A(z,x) = 1$, i.e.~the lowest eigenmode lies exactly
%within the range of the interpolation operators.

We believe that the results shown in this section are quite
encouraging: There is no dependence of the work needed on the lattice
size. Furthermore, except for very high criticality the absolute
number of work units is not too high. For comparison, a
Conjugate Gradient algorithm needs about 3700 work units to solve the
equation with the desired accuracy on the $64^2$-lattice at a mass
value of $10^{-4}$. (Note that the work for the final solution
of the equation is included in figure~\ref{trivlapwork}.) This is about a factor of
two smaller that the work needed by the neural multigrid; however
we have not tuned the parameters to minimize the work;
for instance the number of multigrid sweeps done in each measurement
has been chosen to be~20, which is quite large. In addition, during the
program runs several test measurements of errors, residuals etc.\ are
done that are not strictly necessary. Therefore we believe
that the neural multigrid still has a great potential for improvements.
The real test is of course its behaviour in the case of disordered problems.

\section{Lattice Gauge Theory}\noindent
\subsection{The Problem}\noindent
As an example for a disordered system we consider the
propagator equation for a bosonic particle in an
\SUZWEI-gauge field background\cite{Creutz}. The covariant Laplace-operator in
stencil-notation is 
\begin{equation} \Delta(z) = \left[\matrix{0&U_{z,2}&0\cr
U_{z,-1}&4&U_{z,1}\cr 0&U_{z,-2}&0\cr}
\right]\quad,\label{stencil}\end{equation} 
with $U_{z,\mu} \in {\rm SU(2)}$. The second index denotes the direction of
the coupling to the neighbour.
The link matrices $U_{z,\mu}$ fulfill $U_{z,-\mu} = U_{z-\mu,\mu}^*$.
They are distributed according to the Wilson action
\cite{Wilson}
\begin{equation} S_W = \frac{\beta}{4}
                \sum_{P} {\rm Re\,tr\,}\left(1-U_P \right)
\quad.\end{equation} Here
 $\beta = 4/g^2$ is the inverse coupling and the sum is over all
Plaquettes in the lattice. $U_P$ denotes the parallel transport
around a Plaquette.
This distribution leads to a correlation between
the gauge field matrices with finite correlation length $\chi$
for finite $\beta$. The case
$\beta = 0$ corresponds to a completely random choice of the matrices
($\chi=0$),
for $\beta=\infty$ all matrices are $\matr{1}$
($\chi=\infty$). In this sense, $\beta$
is a disorder parameter, the smaller $\beta$ the shorter the correlation
length and the larger the disorder.

Here we are only interested in the behaviour of our algorithm for a
disordered system and so we will choose in the following $\beta=0$,
which gives the greatest disorder possible.

As it stands, however, the equation is not critical: The lowest
eigenvalue will be quite large (of the order of 0.5). To get a
disordered critical problem we first calculate the lowest eigenvalue
and then subtract it from the diagonal part of the operator. This
allows us to tune criticality and thereby to measure the convergence
behaviour accurately. Note that this destroys the diagonal dominance
of the operator and makes the problem quite difficult for a multigrid
method. 

\subsection{Results}\noindent
Trying the neural multigrid on the described problem, we found quickly
that it was not possible to achieve the desired textbook efficiency of
$\rho=0.1$. A more realistic goal seemed to require the neural
multigrid to reach a $\rho<0.6$, which means that for the same
error reduction we need about four times the work. But if this could
be reached regardless of criticality and lattice size, the algorithm
would still be very efficient.

\begin{figure} \begin{center}
\epsfig{file=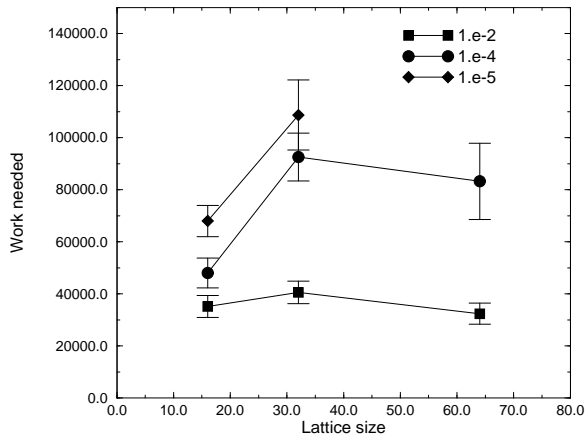,width=8cm}
\end{center}
\caption{Number of Work units needed by the neural multigrid to learn
  interpolation operators with a reduction rate of smaller
  than~0.6 and to solve the \SUZWEI\ Laplace equation. For each data point, ten
  runs were made, except for the $32^2$ and $64^2$ lattices at
  $\dm=10^{-4}$, where 20 and 15 runs were done. \label{gaugelap}} 
\end{figure}

To improve the convergence, we introduced the improvements described
in section~6. However, as a look at
fig.~\ref{gaugelap} shows, the work needed to learn good interpolation
operators that achieve the desired reduction rate grows with the
criticality and at first also with the lattice size. The growth with
the lattice size stops as soon as the $32^2$-lattice is reached, so
for large enough lattices there seems to be no critical slowing down here.
For the growth with the criticality the exponent can be estimated
roughly to be about $z\approx0.4$. Although this is not too bad, the
absolute number of work units is quite large; about 15~times larger
than that needed by a Conjugate Gradient algorithm. 

These results are somewhat disappointing. As it stands, the algorithm
is not as efficient as we hoped for. Nevertheless, we have achieved an
important result: The algorithm was able to find interpolation
operators that allow for multigrid cycles with $\rho<0.6$ in almost
all cases. This means that we have shown numerically that there exists
a practical true multigrid algorithm for a two-dimensional bosonic
gauge theory without critical slowing down. Up to now it was only
known that idealized multigrid algorithms (in four dimensions) were
able to eliminate critical slowing down, but these used non-local
interpolation operators\cite{Thomas}. The ISU algorithm\cite{ISU} also
eliminates critical slowing down for our test problem, but being a
unigrid the work it needs grows as $\ln^2({\rm Volume})$. By
investigating the interpolation operators found by the neural
multigrid more closely we might find valuable informations about how
good interpolation operators should look like. The same can also be
tried for the case of greater interest, namely the four-dimensional
Dirac equation.

\subsection{Overcoming the difficulties}\noindent
Is it possible to overcome the described difficulties?
To answer this, we have to investigate the reason for the problems
with the gauge theory. A first 
hint is that most of the work was done on the coarser layers. This
means that as a twogrid, our algorithm would have (nearly) no critical
slowing down. Are the
coarser layers more problematic than the finest one?

Indeed they are. As the interpolation operators are not simple linear
functions, the effective operator~$\op^k$ has strongly fluctuating
couplings and a fluctuating diagonal term as well. A simple look at
the operators shows that after two blocking steps the coupling
strengths may vary by a factor of ten or more. It is well known that
for couplings with such strong fluctuations a simple blocking
scheme with square blocks will not be efficient.

In order to overcome this difficulty, an algorithm to determine good
shapes for the supports of the interpolation operators is needed. Such
an algorithm was developed as a part of the Algebraic
Multigrid\cite{AMG} and could be used as an ingredient to our neural
multigrid. This algorithm chooses those points as block-centers that
have many strong connections to other points of the lattice and is
quite efficient.

A preliminary study to confirm this picture was also done: We used the
scalar Laplace equation with a site-dependent mass term. At low
criticality, the neural multigrid again exhibited critical slowing
down when the mass term was strongly fluctuating. A study of the
errors showed that indeed these are the points where the error is not
properly reduced.  This problem was partly alleviated by shifting the lattice
such that the points with the weakest connections were not chosen as
block-centers, however some critical slowing down still remained.

So it is probable that using dynamically chosen blocks
the algorithm would perform much better, perhaps even without critical
slowing down. 

\section{Conclusions}\noindent
We have presented a new multigrid method called neural multigrid which
is based on joining multigrid ideas with concepts from neural
nets. The algorithm is able to learn efficient interpolation operators
in the case of the ordered Laplace equation nearly without critical slowing
down and with a surprisingly small amount of work comparable to that
of a Conjugate Gradient solver. 

In the case of a disordered system (the Laplace equation with
$\SUZWEI$ gauge fields) the learning exhibited critical slowing down
with an exponent of about $z\approx0.4$ and the algorithm was able to
find good interpolation operators in this case as well.  Finally it
was shown that the remaining critical slowing down of the algorithm
might be overcome by choosing the supports of the interpolation
operators dynamically.

\nonumsection{Acknowledgments}\noindent
I wish to thank Daniel L\"ubbert for rekindling my interest in neural multigrid.
Financial support by Deutsche Forschungsgemeinschaft is gratefully acknowledged.

\nonumsection{References}

\end{document}